# INFORMATION RETRIEVAL OF JUMBLED WORDS

Venkata Ravinder Paruchuri
Computer Science
Oklahoma State University, Stillwater, OK-74078
venkarp@okstate.edu

## Abstract

It is known that humans can easily read words where the letters have been jumbled in a certain way. This paper examines this problem by associating a distance measure with the jumbling process. Modifications to text were generated according to the Damerau-Levenshtein distance and it was checked if the users are able to read it. Graphical representations of the results are provided.

## Introduction

> *Aoccdrnig to a rscheearch at Cmabrigde Uinervtisy, it deosn't mttaer in waht oredr the ltteers in a wrod are, the olny iprmoetnt tihng is taht the frist and lsat ltteer be at the rghit pclae. The rset can be a toatl mses and you can sitll raed it wouthit porbelm. Tihs is bcuseae the huamn mnid deos not raed ervey lteter by istlef, but the wrod as a wlohe.*

The above text has circulated on the Web for several years to show how powerful the human mind is in making sense of jumbled spellings. It may be viewed from the perspectives of joint error correction and coding [1] that is done simultaneously and automatically by the mind, or from the point of view of approximate string matching [2]-[6].

It has been proposed that the human brain is able to read the words even when they are jumbled because of the following properties

1. The grammatical structure of the sentence is not disturbed in the above sentence, that is the small words [of 2 or 3 letters] or the function words [by, the, is etc] are not jumbled.

Since the grammatical structure is preserved, the user is able to predict the next word in the sentence. The jumbled text not only preserves the grammatical structure, it leaves almost 45-50% of the words correct (In the above paragraph that we took 46% of the words are unchanged.

2. People generally tend to notice the first and last letters more easily than they tend to observe the middle letters. So there is less possibility of finding errors in the middle letters than the initial and last letters.
3. Although the words are jumbled in the paragraph, the jumbled words are not new words, thus making the task of the reader easier.

4. The sound of the original word is preserved in the jumbled words. This also makes reading easy as people tend to read the word by its sound.
5. People read the jumbled text because of the context of the sentence.

The two things that interested me, in this paper, are the use of function words and the context that plays a part in guessing the next word in the sentence. I have decided to remove the function words from the paragraph and then use the same jumbling technique to study the effect of this change. Also, to break the context of the sentence, I have taken 100 independent words that are commonly use in everyday life and then applied the jumbling technique.

## Approximate String Matching

Approximate string matching is the technique of performing string matching to the pattern of text. The match is measured in the number of operations that are performed to match the exact string. The most common operations that are performed to match the string are insertion, deletion and substitution. The number of operations performed is measured in terms of edit distance [13].

Examples of the operations are shown below:

Insertion: monkey → monkeys

Deletion: monkey → money

Substitution: monkey → donkey

All the above operations the number of edit distances performed are one. Some string matchers also consider transposition of two adjacent letters in the string [14].

Transposition: lost →lots

Approximate string matching has applications in many fields. Some examples are recovering the original signals after their transmission over noisy channels, finding DNA subsequences after possible mutations, and text searching where there are typing or spelling errors [6].

Most approximate string matchers assume same cost for all the operations performed in string matching, but some matchers do assign different weights to different operations. A more detailed description about edit distance and distance functions are explained in the distance measures section.

## Distance measures

Edit distance is the number of operations performed to transfer one string into another string. There are different ways of performing the edit distance such as Levenshtein distance [7], Damerau-Levenshtein distance, Hamming distance, Jaro-Winkler distance, Longest common subsequence problem etc.

Levenshtein distance is a metric used to measure the difference between two sequences. This measure between two strings is defined by the number of edit operations used from transforming one string to another. The edit operations may be insertion, deletion and substitution of a single character. Here all the operations cost one unit. Levenshtein distance has a wide range of applications in areas such as spell checkers, dialect pronunciations and used in software's for natural language translations [6].

As example the Levenshtein distance between Sunday and Monday is 2.

Sunday ->Munday (substituting M for S) ->Monday (substituting O for U).

Damerau-Levenshtein distance is similar to Levenshtein distance except that it includes an extra edit operation called the transposition of adjacent letters. Here all the operations also cost one. Damerau-Levenshtein distance has its applications in fields of fraud vendor name detections, where it can detect the letter that has been deleted or substituted, in DNA, where the variation between the two strands of DNA can be found out by this distance [6].

Hamming distance allows only substitution of letters, which cost one unit. It is applied only to the strings of similar length. It is applied in error detection and correction [6].

In this paper we apply the Damerau-Levenshtein distance to the words to find its effect on reading. This is because Damerau-Levenshtein distance has all the possible edit operations that can be performed.

## Experiment and Analysis:

In all the experiments that I have conducted, I recorded the time each of 10 readers took to read the text. This time was then averaged.

### I. Removal of function words

In this section I considered the actual paragraph and then removed all the function words from the paragraph to find the effect on the reader.

**Actual sentence**

> *According to research at an English university, it doesn't matter in what order the letters in a word are, the only important thing is that the first and last letter is at the right place. The rest can be a total mess and you can still read it without problem. This is because we do not read every letter by itself but the word as a whole.*

**After jumbling**

> *Aoccdrnig to a rscheearch at Cmabrigde Uinervtisy, it deosn't mttaer in waht oredr the ltteers in a wrod are, the olny iprmoetnt tihng is taht the frist and lsat ltteer be at the* *rghit pclae. The rset can be a toatl mses and you can sitll raed it wouthit porbelm. Tihs is bcuseae the huamn mnid deos not raed ervey lteter by istlef, but the wrod as a wlohe.*

**Without function words**

*According research English university doesn't matter what order letters word only important thing first last letter right place. Rest total mess still read without problem. This because read every letter itself word whole.*

**After jumbling the above paragraph without function words**

*Accdroing resaecrh elgnsih uvinsreity deosn't mtaetr what order letrets wrod olny iopmrtant tnihg frist lsat lteetr rihgt plcae. rset tatol mses sitll raed whtiuot pborelm. Tihs baceuse raed every lteter istlef word wohle.*

**Results for function words**

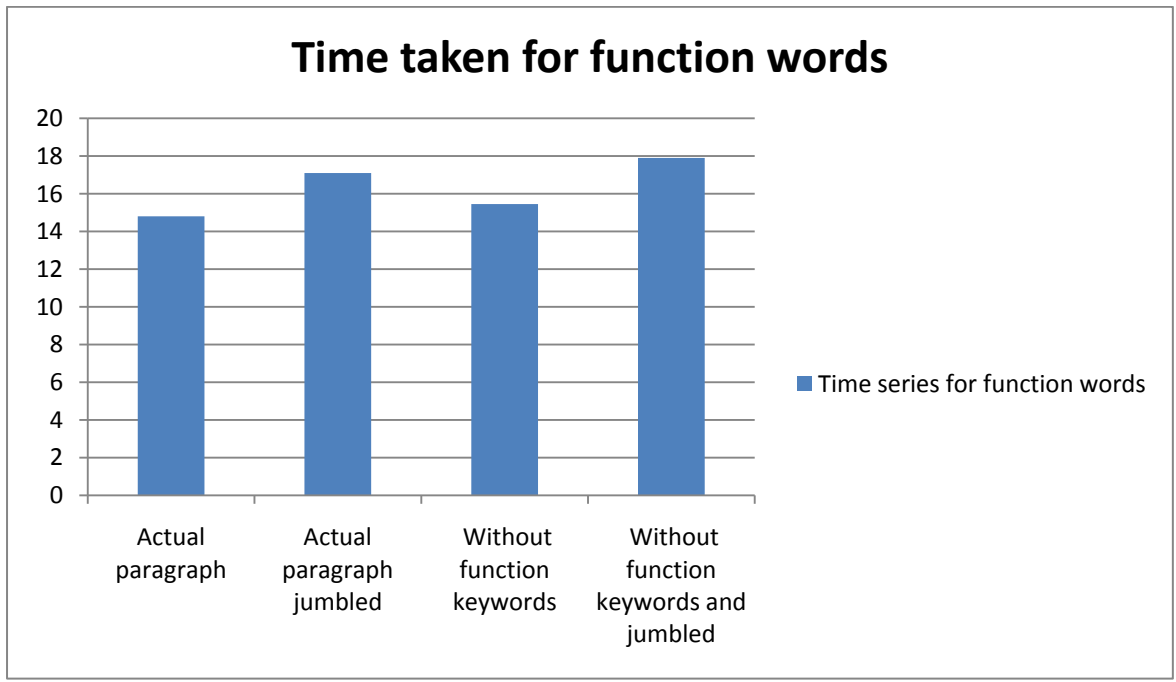

Graph 1: Time taken for function words

From the graph, we can conclude that the function words don't affect the reading.

## II. Jumbling hundred independent words for testing the importance of context

In order to test the importance of context through which the readers are able to read the jumbled sentence, I took some 100 independent words that are commonly used and then jumbled them with their first and last letters in their original position.

**100 independent words considered**

*study basis exciting field utilize great interesting expanding many contributed*

*normal second below genetic graduate notepad jumble compatible giving camping*

*school comment simple scroll action achievement broad paste national essential*

*friend dismal diminish greeting divide coming external proud activate recent*

*money reader search invite competition scientist elevate programs international consistently*

*replace symposium academic followed properties address platform knowledge windows interaction*

*product console fraction participate gaining high people retail average dollar*

*website waste heading several editing potential fragile spending future shoulders*

*burden sector information confused upcoming serious assist substantial quality become*

*common maintain require growing humor animal going finance internet women.*

Then I jumbled these words with their first and last letters in their original position. The resulting words would be:

*sutdy bisas ecxiting feild utlizie gerat inretesitng exapidnng many cnortubietd*

*nomral sceond bolew geneitc grdataue nopetad jbmule coapmbitle giivng cmapnig*

*sohcol cmoemnt sipmle sclorl atcion aihcveeemnt braod psate naitanol esstneial*

*firend dsiaml dimiinsh grteeing divdie conimg exetnral puord aitctave rcenet*

*moeny rdaeer saecrh iivnte cpmotetioin seicntsit eelavte porargms ietnnrtaioanl coisnetsntly*

*rpealce smysopuim aacdimec flloewod pporreites addsers pltarofm konelwgde widnwos itnecartoin*

*porcudt cnolose fcartion paitrpicate ganiing high ppoele rteial aevrgae dollar*

*wsbeite wtsae headnig sevearl editnig ptonetial fargile spennidg fuutre suohedlrs*

*bruedn scetor inofmritaon cfnoused ucpmoing sreuois assist sbusnataitl qlautiy bceome*

*cmmoon matniian reuqrie griwong hmuor ainmal gniog finance ietnnret wemon.*

**The result for independent words**

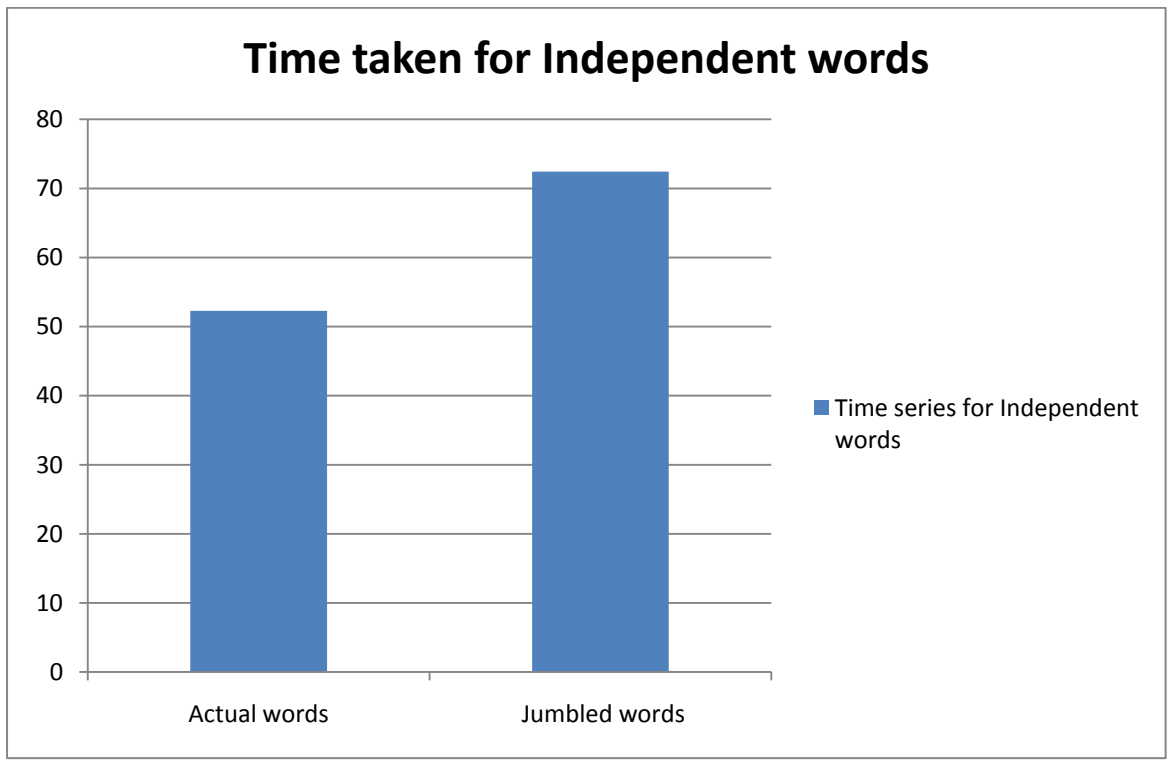

Graph 2: Time taken for independent words

From the above graph, it is clear that the context plays an important role in reading text with jumbled words.

## III. Damerau-Levenshtein distance

I applied the Damerau-Levenshtein distance of one to the original paragraph, obtaining the following text:

> *Accrding to recearch at an Enlgish univiersity, it dosn't matetr in wiat ordier the lettes in a werd are, the only impurtant thng is that the fist and last lette is at the rijght place. The rect can be a totul mess and you can stillt raed it wihout probllem. Tihs is becase we do not raed evry lette by it slf but the wurd as a whule.*

The edit operations performed are addition, deletion, substitution and transposition of neighboring letters. The timings were then recorded.

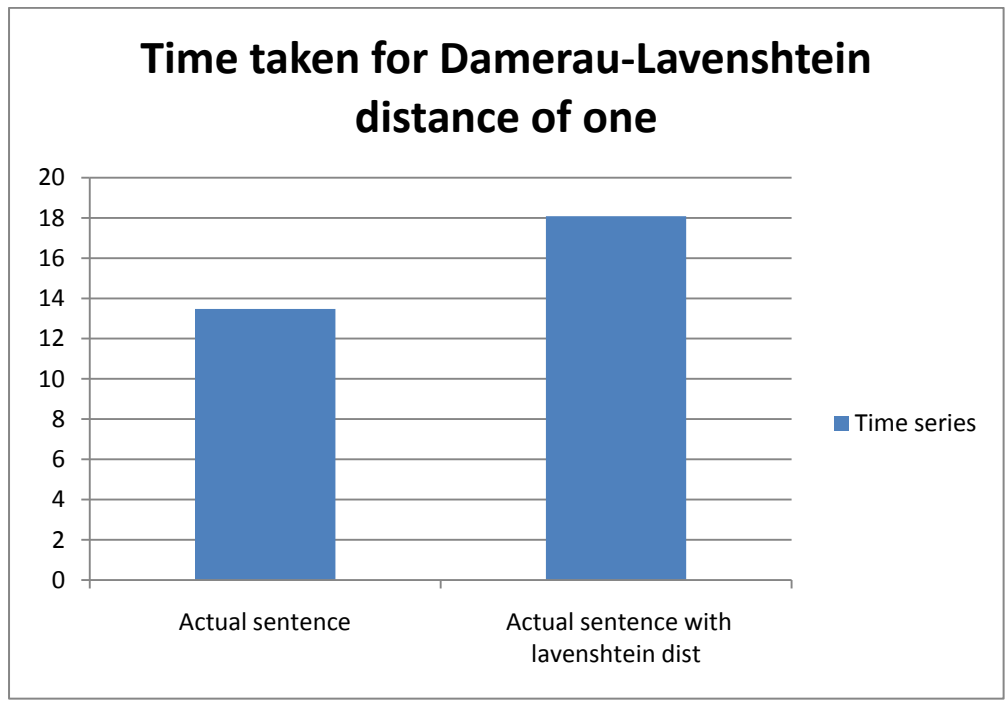

Graph 3: Time taken for Damerau-Lavenshtein distance of one

Then I took the same 100 independent words and used the Damerau-Levenshtein distance of one to find the effect on reading.

On performing the normal Damerau-Levenshtein edit operations on them, the resultant words are:

> *stidy bases exsiting fiel utlize gdeat interesing exapidnng masy contriuted*

>*nirmal sceond elow genietic gracuate nolepad jubmle compaible givin canping*
>
>*schol comaent simpe scrll actiong achievment broud pacte nationale essetnial*
>
>*frind dismul dimnish greating dividee comig extrenal pruud activte recetn*
>
>*monkey rader scarch invitee competetion scientst eelvate progrms internatinal consistnetly*
>
>*reblace sympsium acedemic folowed proerties adrdess platgorm knowlerdge winduws interactiorn*
>
>*produsct consule fractin participte gainnig hicgh peoplie retil averaige dollor*
>
>*wedsite wasteh headng severol editng potemtial frggile speniding futuire shoulers*
>
>*burben sextor infomation confuzed upconing seious asist substtantial guality becme*
>
>*comomn mantain reguire groving humuor anmial goig finunce intrenet wmen*

When the first letter is kept the same and when the edit operations are performed, the resultant words are:

>*sturdy basisd excitng fieldh utilixe greate interseting expading mabny contriduted*
>
>*normasl secod beloe genreic graduyate notead jumbleg comparible givign campingh*
>
>*scxhool coment sinple scrllo actyion achieevment broiad pastre nationale essetnial*
>
>*frind disnal dimiunish greteing dicide cominh extermal proid activatee recebt*
>
>*mney reaser seadch inmvite competotion scientsit elvate progras internatinoal consisrently*
>
>*replece sympsium acaedmic followesd propertiesd addrses platforn knowledghe windwos interactiob*
>
>*producrt consxole fracton participaet gaiing hihg peeple retali avereage dillar*
>
>*webiste wast headign severeal editiong potentila fraglie slending furture shouledrs*
>
>*burcen secror informtion confusef upcomig seriuos asssit substangial quakity becmoe*
>
>*comomn mainhain requier growimg humior aninal goingt finanec internt women*

When the first and last letters are kept the same and then, when the edit operations are performed the resultant words are:

> *Stiudy bacis exiting feild utialize grat intiresting expnading miany contibuted*
>
> *nokmal seocnd beilow gentic gradiate notpead jumible comptible gibing capming*
>
> *schuool coment sinple scrlol acttion achevement broud psate natiional essntial*
>
> *friind disml dimenish gereting dievide comng esternal pruod activhate recnt*
>
> *moniy raeder siearch invte compitition scietnist elievate progams internasional conssitently*
>
> *repilace sympoium acagemic follwoed propehrties addess platfsrm knoweldge wdindows inteaction*
>
> *prxduct conolse fragtion paricipate gaihing hgih peaple reitail avedage dolaar*
>
> *wesbite wagste heding sevaral edtiing poteuntial frgile spebding futrue shoulqders*
>
> *buden segtor informtaion condfused upcming sedious asisst submstantial qulity becone*
>
> *comomn mainytain requre groeing humuor animdal gonig finamce intrnet womwen*

If the edit operations are performed by the neighboring letters in the QWERTY keyboard, the resultant words are:

> *stuidy badis exciitng fielfd utikize grat intwresting ezpanding manuy contrinuted*
>
> *notmal secons belowq generic gradyate notepsd junble compatoble gicing camling*
>
> *schoil commwnt simpke sctoll acyion achiebement beoad psste natonal esesntial*
>
> *friebd didsmal dimnish greerting dicide comiong extrernal produ actuvate rexent*
>
> *mponey reqder seardh invire comletition scietist eleavte proframs intetnational consitently*
>
> *replave symopsium acadrmic follwed proiperties adfress plaform knolwedge windlws ingeraction*
>
> *prouct conaole fratcion partivipate gaiinng hifh peopple reatail avearge dlolar*
>
> *webstie waster hecding seferal edting potetnial frsgile spemding fuyture shulders*
>
> *butden sectpr inforjation confised upcming sreious assitt subsantial qyuality beclme*
>
> *comnon maintin rewuire gruwing hukor animsal goign finanve internrt womwn*

The timings for all the paragraphs were recorded and plotted in Graph 4.

From the graph it can be inferred that if the first and last letters are kept the same and if the edit operations are performed the user is able to read the words with little difficulty. The user is also able to read the words if the edit operations are done with the neighbor letters in the QWERTY keyboard.

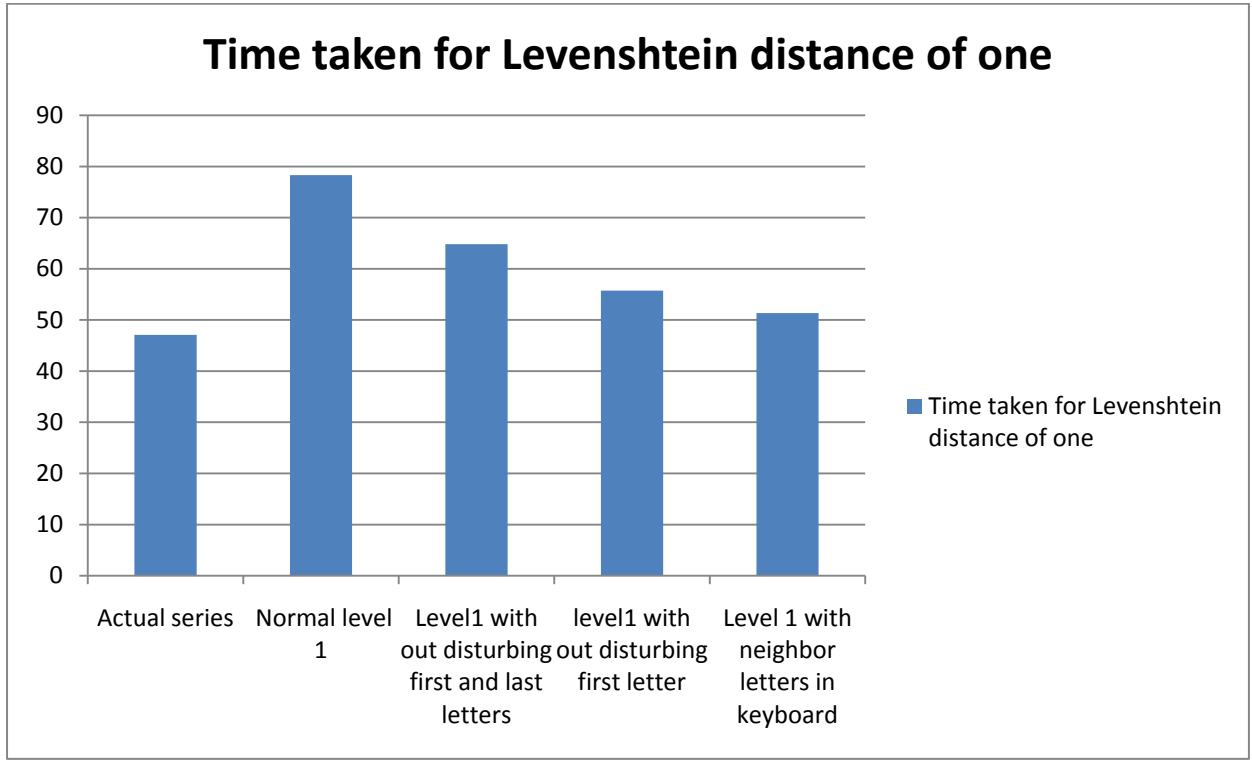

Graph 4: Time taken for Levenshtein distance of one

## Conclusion

This paper tried to explain the idea behind the ease of reading jumbled words. The results show that the importance of functional words in reading these words is much less than proposed before. Nevertheless, context plays an important role in helping the user read such words. In addition, this paper applied the Damerau-Levenshtein distance of one to the words and found that the words can be read if the first and last letters are left in their places.